\documentclass[11pt]{article}
\pdfoutput=1
\usepackage{graphicx,amssymb,amsfonts,amsmath,amssymb,amscd,amstext, mathrsfs}
\makeatletter

\@addtoreset{equation}{section}
\newcommand{\be}{\begin{eqnarray}}
\newcommand{\ee}{\end{eqnarray}}
\newcommand{\ba}{\begin{array}}
\newcommand{\ea}{\end{array}}
\newcommand{\nn}{\nonumber}

\makeatletter \@addtoreset{equation}{section} \makeatother

\begin{document}
~
\vspace{0.5cm}
\begin{center} {\Large \bf  Butterfly Effect and Holographic Mutual Information\\ under External Field and Spatial Noncommutativity}
                                                  
\vspace{1cm}

                      Wung-Hong Huang and  Yi-Hsien Du \\
              ~\\
                       Department of Physics\\
                       National Cheng Kung University\\
                       Tainan, Taiwan\\

\end{center}
\vspace{1cm}
\begin{center} {\large \bf  Abstract} \end{center}
We apply the transformation of mixing azimuthal and internal coordinate or  mixing time and internal coordinate to  a stack of N black M-branes to find  the Melvin spacetime of a stack of  N black  D-branes with magnetic or electric flux  in string theory, after the Kaluza-Klein reduction.  We slightly extend previous formulas to investigate the external magnetic and electric effects on the butterfly effect and holographic mutual information.  It  shows that the Melvin fields do not modify the scrambling time and will enhance  the mutual information.  In addition, we also T-dualize  and twist  a stack of N black D-branes to find a Melvin Universe supported by the flux of the NSNS b-field, which describes a non-comutative spacetime.  It also shows that  the spatial noncommutativity does not modify the scrambling time and will enhance  the mutual information.  We also study the corrected mutual information in the backreaction geometry due to the shock wave in  our three model spacetimes.

\vspace{2cm}
\begin{flushleft}
*E-mail:  whhwung@mail.ncku.edu.tw
\end{flushleft}
\newpage
\tableofcontents
\section {Introduction}
Butterfly effects in the holographic  geometry  have been extensively studied in recent, which led to interesting new insights into quantum chaos and the behaviours of entanglement in near-thermal systems [1,2,3]. These investigations use eternal AdS black hole [4] which has two asymptotically AdS to describe the thermofield double state.   After adding a small perturbation  to one of the two CFTs at early time $t_w$ one then studies the effect on the structure of the state at t = 0.  In this picture  the perturbation can  be modeled by a shock wave near the horizon of the black hole.   In the dual holographic geometry the shock wave will travel across the horizon of the black hole and the mutual information of two space-like regions in each boundary  given by
\be
I(A,B) =S(A)+S(B)-S(A\cup B)
\ee
can be calculated  by the Ryu-Takayanagi prescription for for entanglement entropy [5,6,7].  After an amount of time $t_*$ the mutual information between the two sides is disrupted and it shows the butterfly effect  familiar from chaotic dynamics. 

In the initial investigation [1] the dual black hole geometry in is the non-rotating BTZ black hole.  The later  analysis  has generalized to $AdS_d$ [8], multiple shock waves [2], and localized shock waves
[9,10]. The string corrections of the scrambling time were presented in [11].  The butterfly effect for black holes with rotation or charge had also been studied in [8,12,13].   That on black $D_p$ branes  was investigated in [14].   

In this article, our goals  are to understand how the  external field and spatial non-commutativity will affect the  butterfly effect from the holographic dual geometry.   The previous studies [8,12,13,15] which used the AdS-Reissner-Nordstrom geometry as the model spacetime is not the external field. The EM fields therein are used to form the  black hole itself and dual to the field theory in the present of chemical potential.  In our model the extra fields are the external electric field,  magnetic field  or NSNS b-field,  which are called as Melvin fields [16].   The Melvin electric or magnetic field appears in this paper is the genuinely external field while the NSNS b-field describes the spatial non-commutativity. 

 The stringy effects in scrambling was studied by Shenker and Stanford [11].  They found that the stringy and Planckian correction will increase the scrambling time by
\be
t_*={\beta\over 2\pi}\Big[1+{d(d-1)\ell_s^2\over 4\ell^2_{AdS}}+...\Big]  \log S
\ee  
where $d$ is the space-time dimension of the boundary theory, $\ell_s$ and  $\ell_{AdS}$ are the string and AdS space length respectively. $S$ is the entropy.  In our studies the three Melvin fields, however, do not modify the scrambling time.   This property seems a little strange at first sight. Since the three spacetimes are constructed through the processes of dimensional reduction, twist and/or T-duality it is natural to conjecture that the scrambling time is invariant under any combination of  the three processes.  The general formula in [1] or (2.17) tells us that the scrambling time depends on entropy and temperature.  While it has been known, for a long time, that certain duality transformations do not affect the temperature and entropy of various gravity solutions [17] our three Melvin-field deformed spacetimes provide the interesting examples which have  the same scrambling time. 
       
   The mutual information is an important concept in information theory and is a useful quantity to describe the chaos.  Previous studies on AdS-Reissner-Nordstrom geometry had found that the chemical potential therein (will is proportion to the EM field) will decrease the  mutual information [8,12,13,16].  However,  in our  model spacetime the extra EM fields will increase the  mutual information.  We also find that the  mutual information increases with the increase of the spacial noncommutativity. While the property is consistent to previous literature [18] we furthermore investigate the backreaction property by considering the  mutual information  on the shockwave Melvin field geometry. 

This paper is organized as follows.   In section 2 we first follow the method in  [1] to derive the formula of scrambling time in the more general black branes background and apply it to our model spacetime.  In section 3 we first apply the transformation of mixing azimuthal and internal coordinate [19,20] or mixing time and internal coordinate [21,22] to the 11D M-theory with a stack N black M5-branes [23,24]  to find  the spacetime of a stack of  N black  D4-branes with magnetic or electric flux  in 10 D IIA string theory, after the Kaluza-Klein reduction\footnote{ The method had been applied by us in a previous paper to the zero temperature of M2-branes [25].}. Next, we follow [26]  to begin with a  N black $D_3$ branes and by applying the three chains of operation to find a model spacetime : 1. T-dualize along z to obtain a N black $D_2$ branes [27].  2. “Twist” the compactification.  3. T-dualize along z to obtain Melvin Universe supported by the flux of the NSNS b-field.   After  applying  the mapping of Seiberg and Witten [28]  we then find a new Melvin Universe which describe the non-comutative spacetime  \footnote{The original method  in [23] was applied to zero-temperature spacetime.}.  We calculate the scrambling time in the above three spacetimes to see how the background field will affect it.  

    In section 4  we calculate the  mutual information in the above three spacetimes and determine the  critial interval therein.    In section 5  we  follow the method in  [1,8] to  derive the formula of extremal surface  in the more general black branes background and apply it to study the corrected mutual information in the backreaction geometry due to the shock wave in  our three model spacetimes in section 6.  Last section is used to summarize our results and makes some discussions.
\section {Scrambling Time in General Black Branes Background}
\subsection {Shockwave Geometry}
We first derive the formula of scrambling time in the following  general black branes background
\be
ds^2=-a(r)f(r)dt^2+{dr^2\over b(r)f(r)}+d\Sigma^2_{d-1}
\ee
in which the horizon locates at $r=r_H$ and $f(r_H)=0$ while $a(r_H)\ne0,~b(r_H)\ne0$.  Contract to [1] above metric  has two extra functions $a(r)$ and $b(r)$ which could help us to explicitly see how the external filed will affect the butterfly effect.  The associated temperature is
\be
T={f\rq{}(r_H)\sqrt {a(r_H)b(r_H)}\over 4\pi}
\ee
The line element can be expressed in the Kruskal coordinate
\be
ds^2&=& {4a(r)f(r)\over a(r_H)b(r_H)f\rq{}(r_H)^2}~e^{-f\rq{}(r_H)r_*\sqrt {a(r_H)b(r_H)}}dUdV\\
U&=&e^{{f\rq{}(r_H)\over 2}(-t+r_*)\sqrt {a(r_H)b(r_H)}}\\
V&=&e^{{f\rq{}(r_H)\over 2}(t+r_*)\sqrt {a(r_H)b(r_H)}}
\ee
in which $r_*$ is the tortoise coordinate defined by $dr_*={dr\over f(r) \sqrt {a(r)b(r)}}$. 

 We now follow [1] to add a small null perturbation of asymptotic energy $E\ll M$  (M is the ADM mass of black branes) at  time $t_w$ and radius $r=\Lambda$ in the left asymptotic region.   We denote $\tilde U,~\tilde V$ as coordinates to the left of the perturbation and $U,~V$ are those to the right.  Then, the shell propagating on the constant $U$ surface is described by
\be
\tilde U_w&=&e^{{f\rq{}(\tilde r_H)\over 2}(-t_w+\tilde r_*(\Lambda))\sqrt {a(\tilde r_H)b(\tilde r_H)}}\\
U_w&=&e^{{f\rq{}(r_H)\over 2}(-t_w+r_*(\Lambda))\sqrt {a(r_H)b(r_H)}}
\ee
We can patch two side of geometry and the matching condition which relates $\tilde V$ to $V$  is described by
\be
\tilde U_w \tilde V&=&-e^{{f\rq{}(\tilde r_H)}\tilde r_*(r)\sqrt {a(\tilde r_H)b(\tilde r_H)}}\\
U_w  V&=&-e^{{f\rq{}( r_H)}r_*(r)\sqrt {a( r_H)b( r_H)}}
\ee
In the case of linear order in small $E$ and large $t_w$ the matching condition can be used to find the shift 
\be
\tilde V=V+\alpha
\ee
which is shown in figure 1. 
\\
\\
\scalebox{0.8}{\hspace{5cm}\includegraphics{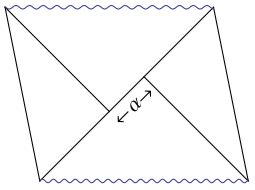}}
\\          
{\it Figure 1:  Penrose diagram of an eternal black hole perturbed by a shock wave.}
\subsection {Scrambling Time}
Since the large value of $t_w$  will lead $r\rightarrow r_H$ we can make a approximation $f(r)\approx f\rq{}(r_H)(r-r_H)+\cdot\cdot\cdot$ and  tortoise coordinate becomes  
\be
r_*(r)\approx\int_0^r{dr\over f\rq{}(r_H)(r-r_H)\sqrt {a(r_H)b(r_H)}}={1\over f\rq{}(r_H)\sqrt {a(r_H)b(r_H)}}\ln\Big((r-r_H)/r_H\Big)
\ee
Therefore
\be
\tilde U_w \tilde V-UV&\approx&-{dr_H\over r_H dM}E
\ee
 In small $E$ we can approx $\tilde U_w\approx U$ [1] and 
\be
\alpha = -{1\over U_w}{dr_H\over r_H dM}E
\ee 

To proceed we will express the shift value in terms of entropy.  Using the black hole area law and thermodynamic first law we can find that
\be
1={TdS_{BH}\over dM}={1\over 4}{TdV_{d-1}(r_H)\over dM}=TS_{BH}{V_{d-1}\rq{}(r_H)\over V_{d-1}(r_H)}{dr_H\over dM}
\ee
and thus
\be
\alpha &=& -{1\over U_w}{E\over TSr_H}{V_{d-1}(r_H)\over V_{d-1}\rq{}(r_H)}\\
&=&-e^{-{f\rq{}(r_H)\over 2}(-t_w+r_*(\Lambda))\sqrt {a(r_H)b(r_H)}}\Big[{E\over T~r_HS_{BH}}{V_{d-1}(r_H)\over V_{d-1}\rq{}(r_H)}\Big]
\ee
The scrambling time $t_*$  is defined to be the value of $t_w$ when $\alpha =1$ and $E\approx T$.  Then we finally find the formula
\be
t_* =r_*(\Lambda)+{\beta \over 2\pi}\ln \Big[{V_{d-1}\rq{}(r_H)\over V_{d-1}(r_H)}{r_HS_{BH}} \Big]
\ee
It is interesting to see that while our general metric is different from that in [1,14] (in which $a(r)=b(r)=1$) the formula of scrambling time $t_*$ has  same form.  Notice that, in the general metric the value of  $V_{d-1}(r_H), ~T,~ S_{BH}$ may be different form that in the case of $a(r)=b(r)=1$ and the scrambling time $t_*$ therein could be different therefore.
\section{Scrambling Time in  Melvin  Field Deformed Geometry }
\subsection{Scrambling Time in  Electric Field Deformed Geometry }
We now apply the transformation of mixing time and internal coordinate to the 11D  M-theory with a stack $N$ black M5 branes to find the spacetime of a stack of  $N$ black D4 branes with electric flux after the Kaluza-Klein reduction. 

The full $N$ black M5-branes solution is given by
\be
ds^2_{11}&=&H^{-1\over3}\left(-f(r)dt^2 +dx_1^2 +dx_2^2+dx_3^2 +dx_4^2 +dx_5^2\right) + H^{2\over3} \left({dr^2\over f(r)}+r^2 d\Omega_4^2\right)\\
d\Omega_4^2 &\equiv& d\gamma^2+cos^2\gamma d\varphi_1^2+sin^2\gamma(d\psi^2 + cos^2\psi\ d\varphi_2^2)
\ee
where $H$ is the harmonic function defined by 
\be 
H = 1+ {R^3\over r^3},~~~R^3 \equiv {16\pi G_{11}\,T_5 \,N\over 3}
\ee
in which  $G_{11}$ is the D-dimensional Newton's constant and $T_5$ is the $M_5$ brane tension.  The function $f(r)$ specified by the horizon at $r_H$ is 
\be
f(r)=1- {r_H^3\over r^3}
\ee

 Now we transform the time $t$ by mixing it with the compactified coordinate $x_5$ in the following substituting 
\be
 t  \rightarrow t - E~ x_{5}
\ee 
Using above substitution the line element (3.1)  becomes
\be
ds^2_{11}&=&-{H^{-1\over3}f(r) \over 1- E^2 f(r)}~dt^2 +H^{-1\over3}\left(dx_1^2 +dx_2^2+dx_3^2 +dx_4^2\right)+ H^{2\over3} \left({dr^2\over f(r)}+r^2 d\Omega_4^2\right)\nn\\
&&~ +H^{-1\over3}\left(1- E^2 f(r)\right) \left(dx_5 +{Ef(r)\over 1- E^2 f(r)}dt \right)^2
\ee
Using the relation between the 11D M-theory metric and string frame metric, dilaton field and 1-form potential
\be
ds_{11}^2= e^{-2\phi/3}ds_{10}^2+  e^{4\phi/3} (dx_{11}+2 A_\mu dx^\mu )^2 
\ee
the 10D IIA background is described by
\be
ds_{10}^2 &=&-{H^{-1\over2}f(r) \over \sqrt {1- E^2 f(r)}}\,dt^2 + \sqrt {1- E^2 f(r)}\left[H^{-1\over2}\left(dx_1^2+dx_2^2+dx_3^2 +dx_4^2\right)+H^{1\over2} \left({dr^2\over f(r)}+r^2 d\Omega_4^2\right)\right]\nn\\
\\
e^{4\Phi\over 3}&=&H^{-1\over3}\left(1- E^2 f(r)\right), ~~~~~ A_{t} ={Ef(r)\over 2(1- E^2 f(r))}
\ee
In this decomposition into ten-dimensional fields which do not depend on the $x_5$, the ten-dimensional Lagrangian density becomes 
\be
L = R- 2 (\nabla \Phi)^2 - e^{2\sqrt 3 \Phi}~ F_{\mu\nu}F^{\mu\nu}
\ee
and from (3.10) we see that the parameter $E$ represents the magnitude of the external electric field.  In the case of  $E=0$  the above spacetime becomes the well-known geometry of a stack of N black D4-branes.  Thus, the background describes the spacetime of a stack of  N black D4-branes with electric flux.

To calculate the thermal quantities of the black branes system we have to change the metric in string frame into the Einstein frame by the relation
\be
g_{\mu\nu}^E=e^{-{\Phi\over 2}}g_{\mu\nu}^{S}
\ee
Thus the line element in Einstein frame is
\be
ds_{E}^2&=&-H^{-3\over8}f(r) (1- E^2 f(r))^{-{7\over8}}\,dt^2 + (1- E^2 f(r))^{1\over8} \Big[H^{-3\over8}\left(dx_1^2+dx_2^2+dx_3^2 +dx_4^2\right)\nn\\
&&+H^{5\over8} \left({dr^2\over f(r)}+r^2 d\Omega_4^2\right)\Big]\nn\\
\ee
In the ``near horizon\rq{}\rq{} limit we  can approximate $H\sim {R^3\over r^3}$ and metric become
\be
ds_{E}^2&=&-\Big({R^3\over r^3}\Big)^{-3\over8}f(r) (1- E^2 f(r))^{-{7\over8}}\,dt^2 + (1- E^2 f(r))^{1\over8} \Big[\Big({R^3\over r^3}\Big)^{-3\over8}\left(dx_1^2+dx_2^2+dx_3^2 +dx_4^2\right)\nn\\
&&+\Big({R^3\over r^3}\Big)^{5\over8} \left({dr^2\over f(r)}+r^2 d\Omega_4^2\right)\Big]
\ee
Use above metric we can calculate the black brane temperature 
\be
T(E)=T_0
\ee
where $T_0$ is that without electirc field  and  scrambling time 
\be
t_*(E)=(t_*)_0
\ee
in which $(t_*)_0$ is the scrambling time  without electric field.  The trivial property is the consequence of  $E^2f(r_H)=0$.
\subsection {Scrambling Time in Magnetic Field Deformed Geometry}
We now  apply the transformation of mixing azimuthal and internal coordinate  to the 11D M-theory with a stack $N$ black M5-branes to find the spacetime of a stack of  N black D4 branes with magnetic flux. 

Using the full N black M5-branes metric described in (3.1) we can transform the angle $\varphi_1$ by mixing it with the compactified coordinate $x_{5}$ in the following substituting
\be
\varphi_1 \rightarrow \varphi_1 + B x_5
\ee
Using the above substitution the line element (3.1)  becomes
\be
\nn ds^{2}_{11}&=&\frac {H^{\frac {2}{3}}r^{2}\cos^{2}\gamma }{1+B^{2}Hr^{2}\cos^{2}\gamma }d\varphi ^{2}_{1}+H^{-\frac {1}{3}}\left( -f\left( r\right) dt^{2}+dx^{2}_{1}+dx^{2}_{2}+dx^{2}_{3}+dx^{2}_{4}\right)
\\
\nn &+&H^{\frac{2}{3}} \left[\frac{dr^{2}}{f \left( r \right) }+r^{2} \left[ d\gamma^{2}+ \sin^{2} \gamma \left( d\psi^{2}+\cos^{2}\psi d\varphi^{2}_{2} \right) \right] \right]
\\
&+&H^{-\frac{1}{3}} \left( 1+B^{2}Hr^{2} \cos^{2} \gamma\right) \left( dx_{5}+ \frac{BHr^{2}\cos ^{2}\gamma }{1+B^{2}Hr^{2}\cos ^{2}\gamma }d\varphi_{1}\right) ^{2}
\ee
Using the relation between the 11D M-theory metric and string frame metric, dilaton field and 1-form potential, the 10D IIA background is then described by
\be
\nn ds^{2}_{10} &=& \frac{H^{\frac {1}{2}}r^{2}\cos ^{2}\gamma}{\sqrt {1+B^{2}Hr^{2}\cos^{2}\gamma }}d \varphi ^{2}_{1}+\sqrt{1+B^{2}Hr^{2}\cos^{2}\gamma }\{H^{-\frac{1}{2}}\left(-f\left( r\right)dt^{2}+dx^{2}_{1}+dx^{2}_{2}+dx^{2}_{3}+dx^{2}_{4}\right)
\\
 &+&H^{\frac{1}{2}}\left[\frac{dr^{2}}{f\left( r\right) }+r^{2}(d\gamma ^{2}+\sin^{2}\gamma\left( d\psi ^{2}+\cos ^{2}\psi d\varphi^{2}_{2}\right) \right]\}
\\
e^{\frac{4}{3}\Phi } &=& H^{-\frac{1}{3}}\left( 1+B^{2}Hr^{2}\cos^{2}\gamma\right), ~~~~~ A^{}_{\varphi^{}_{1}} = \frac {BHr^{2}\cos ^{2}\gamma }{2\left( 1+B^{2}Hr^{2}\cos^{2}\gamma \right) }
\ee
In this decomposition into ten-dimensional fields which do not depend on the $x_5$, the ten-dimensional Lagrangian density will be described by (3.10)  and the parameter $B$ is the magnetic field defined by $B^2=\frac{1}{2}F_{\mu\nu}F^{\mu\nu}|_{\rho=0}$.   In the case of  $B=0$  the above spacetime becomes the well-known geometry of a stack of N black D4-branes.  Thus, the background describes the spacetime of a stack of  N black D4-branes with magnetic flux.

The line element in Einstein frame is
\be
\nn ds^{2}_{E} &=& H^{\frac{5}{8}}r^{2}\cos^{2}\gamma\left( 1+B^{2}Hr^{2}\cos ^{2}\gamma\right) ^{-\frac {7}{8}}d\varphi^{2}_{1}
\\
\nn &+& \left( 1+B^{2}Hr^{2}\cos^{2}\gamma\right)^{\frac{1}{8}} \{ H^{-\frac {3}{8}}\left(-f\left( r\right)dt^{2}+dx^{2}_{1}+dx^{2}_{2}+dx^{2}_{3}+dx^{2}_{4}\right)
\\
 &+& H^{\frac {5}{8}}\left[ \frac {dr^{2}}{f\left( r\right) }+r^{2}\left( d\gamma^{2}+\sin^{2}\gamma\left( d\psi^{2}+\cos ^{2}\psi d\varphi^{2}_{2}\right)\right)\right] \}
\ee
In the ``near horizon\rq{}\rq{} the metric become
\be
\nn ds^{2}_{E} &=& \left( \frac{R^{3}}{r^{3}}\right)^{\frac{5}{8}}r^{2}\cos^{2}\gamma\left( 1+B^{2}\left(\frac {R^{3}}{r^{3}}\right)r^{2}\cos^{2}\gamma\right)^{-\frac{7}{8}}d\varphi^{2}_{1}
\\
\nn &+& \left( 1+B^{2}\left( \frac {R^{3}}{r^{3}}\right)r^{2}\cos^{2}\gamma\right)^{\frac{1}{8}} \Big\{ \left( \frac {R^{3}}{r^{3}}\right)^{-\frac {3}{8}}\left(-f\left(r\right)dt^ {2}+dx^{2}_{1}+dx^{2}_{2}+dx^{2}_{3}+dx^{2}_{4}\right)
\\
 &+& \left( \frac {R^{3}}{r^{3}}\right)^{\frac{5}{8}}\left[\frac {dr^{2}}{f\left(r\right) }+r^{2}\left( d\gamma^{2}+\sin^{2}\gamma \left(d\psi^{2}+\cos^{2}\psi d\varphi^{2}_{2}\right) \right) \right] \Big\}
\ee
Use above metric we can calculate the black brane temperature 
\be
T(B)=T_0
\ee
where $T_0$ is that without magnetic field  and  scrambling time 
\be
t_*(B)=(t_*)_0 
\ee
in which $(t_*)_0$ is the scrambling time  without magnetic field.  Therefore, likes as  the Malvin electic field the Malvin magnetic field also does not modify the scrambling time.
\subsection{Scrambling Time in  Noncommutative Geometry}
To begin with, we quote the formula of T-duality [27].  After the T-duality on $z$ coordinate the metric, dilaton field  and B field become: 
\be
\tilde g_{zz} &=&1/g_{zz}, ~~~\tilde g_{\mu z} = B_{\mu z}/ g_{zz},~~~\tilde g_{\mu \nu} = g_{\mu \nu}-(g_{\mu z}g_{\nu z} - B_{\mu z}B_{\nu z})/g_{zz}\\
e^{-2\tilde \phi} &=& g_{zz}~e^{-2\phi},~~~\tilde B_{z\mu } = g_{z\mu }/ g_{zz},
\ee
Now consider the metric of a stack of  $N$ black $D_3$ branes 
\be
 ds^{2}_{10}&=&H^{-1/2}\Big[-f(r) dt^2+d\rho^2+\rho^2 d\phi^2+d z^2\Big]+H^{1/2}\Big[f(r)^{-1} dr^2+r^2d\Omega_5^2\Big]
\ee
which has zero dilation field and $f(r)=1-r_h^4/r^4$.   T-dualize along $ z$  gives  a stack of  $N$ black $D_2$ branes which is  described by
\be
 ds^{2}_{10}&=&H^{-1/2}\Big[-f(r) dt^2+d\rho^2+\rho^2 d\phi^2\Big]+H^{1/2}\Big[d\tilde z^2+f(r)^{-1} dt^2+r^2d\Omega_5^2\Big]\\
 e^{-2\Phi}&=&H^{-1/2}
\ee
After performing the  twist  by replacing
\be
\phi\rightarrow \phi+ b \tilde z
\ee
and then T-dualize along $ \tilde z$  will give  a stack of  $N$ black $D_3$ branes described by
\be
 ds^{2}_{10}&=&H^{-1/2}\Big[-f(r) dt^2+d\rho^2+{\rho^2 d\phi^2+dz^2\over 1+H^{-1}b^2\rho^2}\Big]+H^{1/2}\Big[f(r)^{-1} dt^2+r^2d\Omega_5^2\Big]\\
 e^{-2\Phi}&=&1+H^{-1}b^2\rho^2\\
 B_{z\phi}&=&{H^{-1}b\rho^2\over 1+H^{-1}b^2\rho^2}
\ee
in which  $ b_{z\phi}$  is the flux of the NSNS b-field.   After  applying  the mapping of Seiberg and Witten [25] 
\be
(G+\theta)^{\mu\nu}=[(g+b)_{\mu\nu}]^{-1}
\ee
the metric $G$ and non-commutativity parameter $θ$ are
\be
G_{\mu\nu}dx^\mu dx^\nu&=&H^{-1/2}\Big[-f(r)d\rho^2+{\rho^2 d\phi^2+dz^2}\Big] +H^{1/2} \Big[f(r)^{-1} dr^2+r^2d\Omega_5^2\Big]\\
\theta^{z\phi}&=&b
\ee
which describes the black branes geometry with non-commutativity $\theta^{z\phi}=b$ [23].  Note that in Cartesian coordinates the non-commutativity parameter is non-constant.

The line element in Einstein frame is
\be
 ds^{2}_{10}&=&\Big(1+{r^4\over R^4}\rho^2 B^2\Big)^{1\over 4}\Big[\Big({R^4\over r^4}\Big)^{-1\over 2}\Big(-f(r) dt^2+d\rho^2+{\rho^2 d\phi^2+dz^2\over 1+\Big({R^4\over r^4}\Big)^{-1}B^2\rho^2}\Big)\nn\\
 &&+\Big({R^4\over r^4}\Big)^{1\over2}\Big(f(r)^{-1} dr^2+r^2d\Omega_5^2\Big)\Big]
\ee
in which the ``'near horizon'  lime  $H\rightarrow {R^4\over r^4}$ is used.

Use above metric we can calculate the black brane temperature 
\be
T(B)=T_0
\ee
where $T_0$ is that without NSNS B-field  and  scrambling time 
\be
t_*(B)=(t_*)_0
\ee
in which $(t_*)_0$ is the scrambling time  without  NSNS B-field.  The trivial property is a consequence of  the invariance of entropy under T duality [18].
\section {Mutual Information in Melvin Field Deformed Geometry}
In this section we will compute the mutual information of a region A on the left asymptotic
boundary and its partner B on the right asymptotic boundary. Note that A = B when the left and right boundaries are identified. For simplicity we consider  the mutual information (1.1) between two strips  contained in the left and right side of the geometry respectively. Therefore we can use the Ryu-Takayanagi prescription  [5,6,7] to calculate the entangle entropy (EE) between each strip and the rest of the system ($S(A$) and $S(B)$), and the EE of their union ($S(A\cup B)$).  \\

To  consider a strip we can pick up an interval $L$ in a coordinate $y$ with 
\be
-{L\over 2}<y<{L\over 2}
\ee
Since that in the black brane system there is a function $f(r)$ which is zero on horizon $r=r_H$ we can let $y=y(r)$ in the Ryu-Takayanagi prescription and, after integrating other coordinates besides $r$,  the action  can be described by 
\be
{\cal A}=2\int_{r_{min}}^\infty dr~ C(r)\sqrt {f(r)^{-1}+D(r) y\rq{}(r)^2}
\ee
in which $r_{min}$ is the turning point.  This is the general action which includes those in [8,14] and  mode spacetimes studied in this paper.  A minimal bulk surface is found by extremzing the area functional, which  leads to
\be
 y\rq{}(r) ={1\over \sqrt {f(r)D(r)\Big({C^2(r)D(r)\over C^2_{min}D_{min}}-1\Big)}}
\ee
in which $C_{min}\equiv C(r_{min})$ and $D_{min}\equiv D(r_{min})$. Note that the interval $L$ can be calculated by
\be
L=2 \int_{r_{min}}^\infty dr~ y\rq{}(r)
\ee
 In this way the minimum surface becomes
\be
{\rm Area}= 2\int_{r_{min}}^\infty dr~ {C(r)\over \sqrt{f(r)}}{1\over \sqrt {1-{C^2_{min}D_{min}\over C^2(r)D(r)}}} 
\ee
and 
\be
I(L)=S(A)+S(B)-S(A\cup B)= \int_{r_{min}}^\infty dr~ {C(r)\over \sqrt{f(r)}}{1\over \sqrt {1-{C^2_{min}D_{min}\over C^2(r)D(r)}}} -\int_{r_{H}}^\infty dr~ {C(r)\over \sqrt{f(r)}}
\ee
where the second term is $S(A\cup B)$ that from area of surface which passes through the horizon
to connect to the other side [1].

We now want to  find the  critical interval $L_c$ in which the mutual information becomes zero.  To proceed, we first rewrite the mutual information function as
\be
I(L)&=& \sqrt{C^2_{min}D_{min}} ~ {L\over 2} + \int_{r_{min}}^\infty dr~ {C(r)\over \sqrt{f(r)}}{ \sqrt {1-{C^2_{min}D_{min}\over C^2(r)D(r)}}}  - \int_{r_{H}}^\infty dr~ {C(r)\over \sqrt{f(r)}}
\ee
Notice that above result is an exact relation. Since the mutual information becomes zero when $r_{min}\approx r_H$  above relation leads to 
\be
L_c &=&{2\over  \sqrt{C^2_{H}D_{H}}} \int_{r_{H}}^\infty dr~ {C(r)\over \sqrt{f(r)}}\left[1- { \sqrt {1-{C^2_{H}D_{H}\over C^2(r)D(r)}}}\right]\\
&\approx&C(r_H) \sqrt{D(r_H)}\int_{r_{H}}^\infty~dr~{1\over \sqrt{f(r)}~C(r)D(r)} + \cdot\cdot\cdot
\ee
Above formula gives same scaling relation between $L_c$ and $R$ as that in [8].  We now apply these formulas to the Melvin field deformed geometry. 
\subsection{Mutual Information and Critical Interval in Electric Field Deformed geometry}
 For the electric field deformed geometry the action becomes
\be
{\cal A}(E)={16\pi^2\over 3}\int_{r_{min}}^\infty dr~ rR^3~(1-E^2 f(r))^{1\over2} ~\sqrt {f(r)^{-1}+{r^4\over R^4} y\rq{}(r)^2}
\ee 
Using above formulation we present  figure 2 to show the mutual information in Melvin electric field spactime.
\\
\\
\scalebox{0.8}{\hspace{3cm}\includegraphics{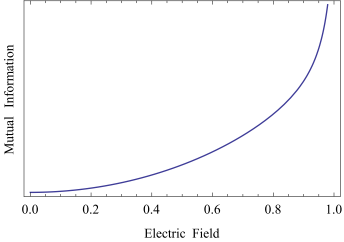}}
\\          
{\it Figure 2:   Mutual information under Melvin electric field for the case of  $r_H=1$ and $R=10$.}
\\
The associated critical interval $L_c(E)$ calculated from above formula is plotted in figure 3.
\\
\\
\scalebox{1}{\hspace{3cm}\includegraphics{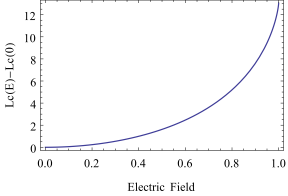}}
\\          
{\it Figure 3:   Critical interval $L_c(E)$ of the mutual information under Melvin electric  field for the case of  $r_H=1$ and $R=10$.}
\\
\\
Use the approximation relation
\be
L_c(E) &\approx&{1\over r_H^3} \int_{r_{H}}^\infty dr~ {1\over{r^5}~\sqrt{f(r)}\sqrt{1-E^2f(r)}}+\cdot\cdot\cdot
\ee
the increasing property of the function $L_c(E)$ can be easily read.
\subsection{Mutual Information and Critical Interval in Magnetic Field Deformed geometry}
 For the small-magnetic field deformed geometry the action becomes
\be
{\cal A}(B)= \int_{r_{min}}^\infty dr~ rR^3~\Big(1+{2B^2R^3\over 5r}\Big))^{1\over2} ~\sqrt {f(r)^{-1}+{r^3\over R^3}\Big(1+{8B^2R^3\over 35r}\Big) y\rq{}(r)^2}
\ee 
Using above formulation we present  figure 4 to show the mutual information in Melvin magnetic field spactime.
\\
\\
\scalebox{0.8}{\hspace{3cm}\includegraphics{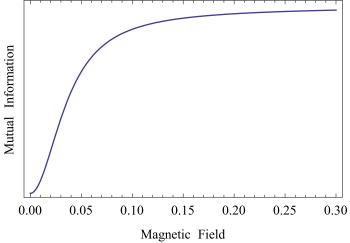}}
\\          
{\it Figure 4:   Mutual information under Melvin electric  field for the case of  $r_H=1$ and $R=10$.}
\\
The associated critical interval $L_c(B)$ calculated from above formula is plotted in figure 5.
\\
\\
\scalebox{0.8}{\hspace{3cm}\includegraphics{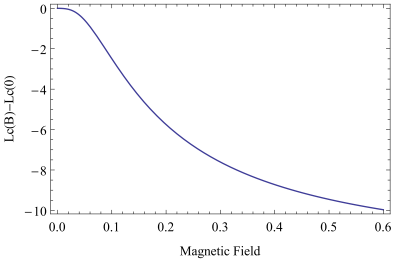}}
\\          
{\it Figure 5: Critical interval $L_c(B)$ of the mutual information under Melvin magnetic  field for the case of  $r_H=1$ and $R=10$.}
\\
\\
Use the approximation relation
\be
L_c(B) &\approx&{1\over r_H^3} \int_{r_{H}}^\infty dr~ {1\over{r^5}~\sqrt{f(r)}\sqrt{\Big(1+{2B^2R^3\over 5r}\Big)}}+\cdot\cdot\cdot
\ee
the decreasing property of the function $L_c(B)$ can be easily read.
\subsection{Mutual Information and Critical Interval in Noncommutative Geometry}
For the NSNS b-field   deformed geometry the action becomes
\be
{\cal A}(b)= \int_{r_{min}}^\infty dr~ r\rho R^4 \sqrt{1+{r^4\over R^4}b^2} ~\sqrt {f(r)^{-1}+{{r^4\over R^4}\over 1+{r^4\over R^4}b^2}~ y\rq{}(r)^2}
\ee 
Using above formulation we present  figure 6 to show the mutual information in Melvin NSNS b-field spactime.
\\
\\
\scalebox{0.8}{\hspace{3cm}\includegraphics{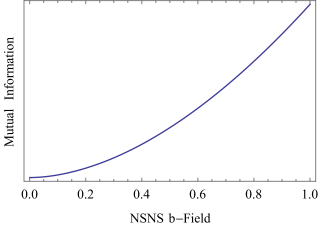}}
\\          
{\it Figure 6:   Mutual information under NSNS b-field for the case of  $r_H=1$ and $R=10$.}
\\
The associated critical interval $L_c(b)$ calculated from above formula is plotted in figure 7.
\\
\\
\scalebox{1}{\hspace{3cm}\includegraphics{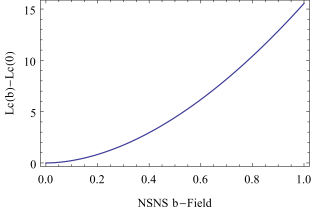}}
\\          
{\it Figure 7:  Critical interval $L_c(b)$ of the mutual information under NSNS b-field for the case of  $r_H=1$ and $R=10$.}
\\
\\
Use the approximation relation
\be
L_c(b) &\approx& \int_{r_{min}}^\infty dr~ \rho ~ { R^3 r_H^3~\sqrt{1+{r^4\over R^4}b^2} \over r^5~\sqrt{f(r)}}+\cdot\cdot\cdot
\ee
the increasing property of the function $L_c(b)$ can be easily read.
\\

Above results show that the Melvin electric field, Melvin magnetic field and NSNS b-field  will increase  the mutual information.  

\section {Extremal Surface in Shockwave Melvin Field Geometry}
We now consider the mutual information in the backreaction geometry due to the present of shock wave.  
\subsection {Extremal Surfaces}
The area of minimal surface is reduced to two-dimensional problem and is dercribed by $r(t)$ [8].  It can be studied from the general form
\be
{\cal A}=\int dt~\sqrt{-a(r)f(r)+{\dot r^2\over b(r)f(r)}}
\ee
Regarding ${\cal A}$ as a particle action we see that the t-translation symmetry in there gives a conserved quantity 
\be
\gamma\equiv \sqrt{-a(r_0)f(r_0)}={-a(r)f(r)\over \sqrt{-a(r)f(r)+{\dot r^2\over b(r)f(r)}}}
\ee
where $r_0$ is define as the radial position where $\dot r=0$.  Since  that  $r_0$ is behind the horizon and thus $f(r_0)$ is negative as shown in figure 8.  Above relation implies that
\be
dt ={dr\over \sqrt{a(r)b(r)f(r)^2 \Big(1+{a(r)f(r)\over \gamma^2}\Big)}}
\ee
\\
\\
\\
\scalebox{0.8}{\hspace{5cm}\includegraphics{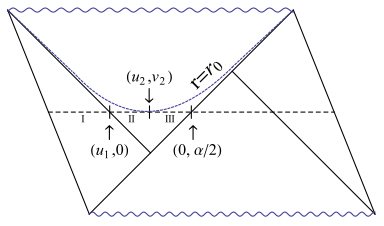}}
\\          
{\it Figure 8:  The Penrose diagram and minimal surface (dashing horizontal line) in the shockwave geometry. The left half of the surface is split into three segments, labeled I, II, and III. The smallest value of r attained by the surface is $r = r_0$, which marks the division between II and III and locates at $(u_2,v_2)$.}
\\
\\
Therefore the entanglement entropy can be evaluated from the extremal surface [5,6,7]  
\be
S_{A\cup B}(r_0)={{\cal A}\over 4}={1\over 4}\int dr~ {\sqrt{a(r)}\over \sqrt{b(r)\Big(\gamma^2+ a(r)f(r)\Big)}}
\ee
In order to examine  how the EE depends on $\alpha$ we have to find the relation between $\alpha$ and $r_0$.  In below, for self-consistent,  we will slightly generalize [8] to derive this relation in our model spacetimes. 
\subsection {Surface Location}
For the segment I, the minimal surface stretches from the boundary at $(u, v) = (1,1)$ to $(u; v) = (u_1; 0)$. Using the definition of $u$ we have 
\be
u_1&\equiv&{\rm exp}\Big[{f'(r_H)\sqrt{a(r_H)b(r_H)}\over 2}\Big(\Delta r_* -\Delta t\Big)\Big]\nn\\
&=&{\rm exp}\left[{2\pi\over \beta}\int^R_\infty {dr\over f(r)\sqrt{a(r)b(r)}}\left(1-{1\over \sqrt{1+{a(r)f(r)\over \gamma^2}}}\right)\right]
\ee
Since that  $r=\infty$ at left boundary and $r=r_H$ at right boundary $(u_1,0)$ as shown in figure 5. 

For the segment II, the $u_2$ can be evaluated in a same way and
\be
u_2&=&{\rm exp}\left[{2\pi\over \beta}\int^{r_0}_\infty {dr\over f(r)\sqrt{a(r)b(r)}}\left(1-{1\over \sqrt{1+{a(r)f(r)\over \gamma^2}}}\right)\right]
\ee
Since that  $r=r_0$ at right boundary $(u_2,v_2)$ as shown in figure 8. We now need to find the formula about $v_2$.  In this case we can define a reference point $(\bar u, \bar v)$ which is in the interior of the black brane and  at some radial coordinate $r = \bar r < r_H$.  Then
\be
v_2&\equiv&{\bar u\bar v\over u_2}{\rm exp}\Big[f\rq{}( r_H)\Delta r_*(r)\sqrt {a( r_H)b( r_H)}\Big]={\bar u\bar v\over u_2}{\rm exp}\left[{4\pi\over \beta}\int^{r_0}_{\bar r} {dr\over f(r)\sqrt{a(r)b(r)}}\right]
\ee
Note that if we choose the point located at $r_*(\bar r)=0$ then $\bar u\bar v=1$ [8].

For the segment III, the  right boundary is at $(u,v)\equiv (u_3,v_3)=(0, {\alpha\over 2})$ as shown in figure 8.  Using the definition of $v$ we have
\be
{v_3\over v_2}&\equiv&{ \alpha\over 2v_2}={\rm exp}\Big[{f'(r_H)\sqrt{a(r_H)b(r_H)}\over 2}\Big(\Delta r^* +\Delta t\Big)\Big]\nn\\
&=&{\rm exp}\left[{2\pi\over \beta}\int^{r_H}_{r_0} {dr\over f(r)\sqrt{a(r)b(r)}}\left(1-{1\over \sqrt{1+{a(r)f(r)\over \gamma^2}}}\right)\right]
\ee
Since that  $r=r_0$ at left boundary $(u_2,v_2)$ and $r=r_H$ at right boundary $(0, {\alpha\over 2})$ as shown in figure 8.

Combine above three relations we can find that
\be
\alpha =2\bar u\bar v~{\rm exp}\left[K_1+K_2+K_3\right]
\ee
where
\be
K_1&=&{4\pi\over \beta}\int^{r_0}_{\bar r} {dr\over f(r)\sqrt{a(r)b(r)}}\\
K_2&=&{2\pi\over \beta}\int_{r_H}^\infty {dr\over f(r)\sqrt{a(r)b(r)}}\left(1-{1\over \sqrt{1+{a(r)f(r)\over \gamma^2}}}\right)\\
K_3&=&{4\pi\over \beta}\int^{r_H}_{r_0} {dr\over f(r)\sqrt{a(r)b(r)}}\left(1-{1\over \sqrt{1+{a(r)f(r)\over \gamma^2}}}\right)
\ee
Using above relation  the entanglement entropy can be expressed as function $\alpha$. Since it is divergent we can renormalized it by subtracting its value at  $\alpha=0$.  Then
\be
S^R_{A\cup B}(\alpha)=S_{A\cup B}(\alpha) - S_{A\cup B}(0)
\ee
In next section we will use above formulas to numerically study the mutual information in three kinds of shockwave Melvin field geometry.  
\section {Mutual Information in Shockwave Melvin Field Geometry}
\subsection {Mutual Information in Shockwave Electric Field Deformed Geometry}
For the  Melvin electric field deformed geometry the minimal surface is described by
\be
{\cal A}&=&{1\over 4}\int dt~\sqrt {R^3\over r^3}~r^4 \sqrt {1-E^2f(r)}~\nn\\
&&\times~\sqrt{-\Big({R^3\over r^3}\Big)^{-3\over8}\Big(1-E^2f(r)\Big)^{-7\over8}f(r)+\Big({R^3\over r^3}\Big)^{5\over8}\Big(1-E^2f(r)\Big)^{1\over8}{\dot r^2\over f(r)}}
\ee
The relation between $\alpha$ and $r_0$ is shown in figure 9. 
\\
\\
\scalebox{0.7}{\hspace{3cm}\includegraphics{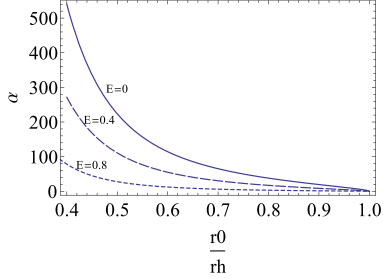}}
\\          
{\it Figure 9:  Relation between $\alpha$ and $r_0$ under Melvin electric field E.}
\\
\\
The figure 10 describes the  renormalized entanglement entropy ``$- S^R_{A\cup B}$'' as function of  Malvin electric  field  $E$.  Note that the total  entanglement entropy is $S(A)+S(B) - S(A\cup B) - S^R_{A\cup B} $ and  thus  ``$ - S^R_{A\cup B}$'' describes the corrected mutual information from backreaction geometry due to the shock wave. 
\\
\\
\scalebox{0.7}{\hspace{3cm}\includegraphics{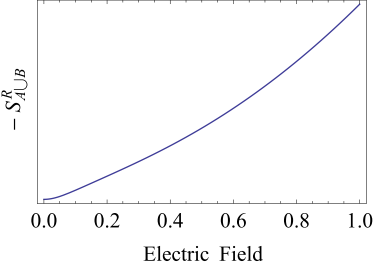}}
\\          
{\it Figure 10:  Renormalized entanglement entropy ``$ - S^R_{A\cup B}$'' as function of  Malvin electric  field  $E$.}
\subsection {Mutual Information in Shockwave Magnetic Field Deformed Geometry}
For the  Melvin magnetic  field deformed geometry the minimal surface is described by
\be
{\cal A}\approx{1\over 4}\int dt~\sqrt {R^3\over r^3}~r^4\Big({1+{B^2R^3/r\over 40}}\Big)~\sqrt{-\Big({R^3\over r^3}\Big)^{-3\over8}f(r)+\Big({R^3\over r^3}\Big)^{5\over8}{\dot r^2\over f(r)}}
\ee
in which we adopt the approximation for the case of small $B$ field. The relation between $\alpha$ and $B$,  and renormalized entanglement entropy $ - S^R_{A\cup B}$ as function of  Malvin magnetic field $B$ are  shown in the figures 11 and 12.
\\
\\
\scalebox{0.7}{\hspace{3cm}\includegraphics{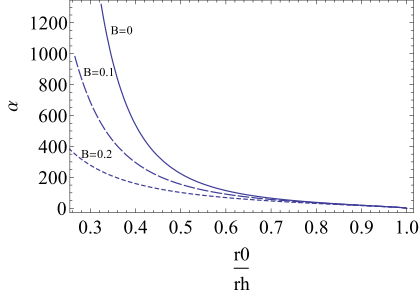}}
\\          
{\it Figure 11:  Relation between $\alpha$ and $r_0$ under Melvin magnetic field B.}
\\
\\
\\
\scalebox{0.7}{\hspace{3cm}\includegraphics{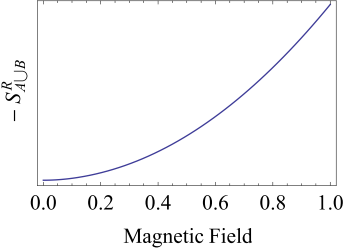}}
\\          
{\it Figure 12:  Renormalized entanglement entropy ``$-S^R_{A\cup B}$'' as function of  Malvin magnetic  field  $B$.}
\subsection {Mutual Information in Shockwave Noncommutative Geometry}  
For the  NSNS b-field deformed geometry the minimal surface is described by
\be
{\cal A}\approx{1\over 4}\int dt~\sqrt {R^4\over r^4}~r^4\Big({1+{b^2r^4/R^4\over 8}}\Big)~\sqrt{-\Big({R^4\over r^4}\Big)^{-1\over2}f(r)+\Big({R^4\over r^4}\Big)^{1\over2}{\dot r^2\over f(r)}}
\ee
in which we adopt the approximation for the case of small $b$ field. The relation between $\alpha$ and $r_0$,  and renormalized entanglement entropy $-~S^R_{A\cup B}$ as function of  non-commutativity $b$ are  shown in the figures 13 and 14.
\\
\\
\scalebox{0.7}{\hspace{3cm}\includegraphics{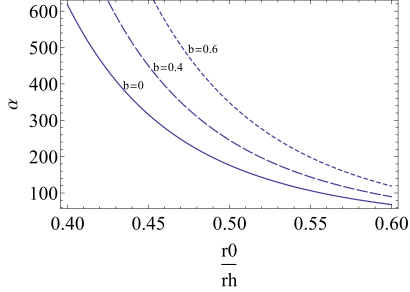}} 
\\          
{\it Figure 13:  Relation between $\alpha$ and $r_0$ under Melvin NSNS b-field.}
\\
\\
\\
\scalebox{0.7}{\hspace{3cm}\includegraphics{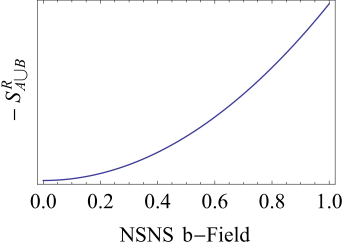}}
\\          
{\it Figure 14:  Renormalized entanglement entropy ``$ - S^R_{A\cup B}$'' as function of  non-commutativity $b$.}
\\
\\
All above figures show that Melvin fields will enhance  the mutual information in the backreaction geometry due to the shock wave. 
\\
\section {Conclusion}
 The  idea that entanglement entropy has some resemblance to thermodynamic entropy including a sort of first law-like relation has been vigorously investigated recently from the field theoretical and from the holographic sides  [29,30]. The study of mutual information might also serve to determine which set of quantities are expected to be different.  In this paper we  extend previous method [25,26] to derive three kinds of Melvin geometry. First and second geometries describe the spacetime of a stack of  N black  D4-branes with magnetic or electric flux  in 10 D IIA string theory. The third geometry describes a Melvin Universe supported by the flux of the NSNS b-field,  which relates  to the non-comutative spacetime. We then follow the method in  [1,8] to derive the new formula of scrambling time and  holographic mutual information in more general spaceimes and use the new formulas to investigate the scrambling time and  holographic mutual information in the three model spacetimes.  We first find that while the three kinds of external field do no modify the scrambling time they can enhance the  mutual information.  We also study the  mutual information corrected by the backreaction geometry due to the shock wave.   It also shows that the three kinds of external field will enhance the mutual information.    
 
  Finally, we shall mention that the prescription of Melvin twist used in this paper  is a powerful solution generating  technique in string theories [19,20,21,22].  Applying it to the Dp-brane background and the subsequent near horizon limit gives rise to supergravity duals for a variety of decoupled field theories depending on the orientation of the brane and the Melvin twist.   There are  many  geometries which can be generated from a slight variation of the Melvin twist, as shown in the table of reference [26].    One of our three geometries describes the supergravity background dual to the non-commutative gauge theory, in which  the non-commutativity are not constant values [26,31,32].  The electric-field deformed and magnetic-field deformed spacetimes are related to the fluxbranes [33,34].  The holographic duality for the fluxbranes remains an interesting and open problem [34].  Note that the procedure of Melvin twist  relies on having a U(1) $\times$ U(1) compact isometry along which one performs a sequence of T-duality, twist, and a T-duality. For examples, if one of the U(1) is transverse to the brane, then one obtains a dipole field theory [35,36,37]. If taking both of the U(1) to be transverse to the brane gives rise to the construction of Lunin and Maldacena [38].  It is interesting to investigate the quantum chaos  on these spacetimes.  It is also interesting  to see that whether the properties found in this paper could be shown in any simple field theory models.  We are now searching the toy mode and investigating the properties from the field theoretic side.
\\
\\
Acknowledge : YHD is supported by the Ministry of Science and Technology of Taiwan (MOST-103-2112-M-006-004-MY3).
\\
\section{References}
\begin{enumerate}
\item S. H. Shenker and D. Stanford, ``Black holes and the butterfly effect,'' JHEP 1403
(2014) 067 [arXiv:1306.0622 [hep-th]].
\item S. H. Shenker and D. Stanford, ``Multiple Shocks,''JHEP 1412 (2014) 046
[arXiv:1312.3296 [hep-th]].
\item J. Maldacena, S. H. Shenker and D. Stanford, ``A bound on chaos,'' [arXiv:1503.01409
[hep-th]].
\item J. M. Maldacena, ``Eternal black holes in anti-de Sitter," JHEP 0304, 021 (2003)
[hep-th/0106112]. 
\item S. Ryu and T. Takayanagi, ``Holographic derivation of entanglement entropy from AdS/CFT,''
Phys. Rev. Lett. 96, 181602 (2006) [hep-th/0603001].
\item S. Ryu and T. Takayanagi, ``Aspects of Holographic Entanglement Entropy,'' JHEP 0608, 045
(2006) [hep-th/0605073].
\item I. R. Klebanov, D. Kutasov and A. Murugan, ``Entanglement as a probe of confinement,'' Nucl.
Phys. B 796, 274 (2008) [arXiv:0709.2140 [hep-th]].
\item S. Leichenauer, ``Disrupting Entanglement of Black Holes,'' Phys. Rev. D90 (2014) 046009,
[[arXiv: 1405.7365 [hep-th]].
\item D. A. Roberts, D. Stanford and L. Susskind, ``Localized shocks,'' JHEP 1503 (2015)
051 [arXiv:1409.8180 [hep-th]].
\item P. Caputa, J. Simón, A. Štikonas, T. Takayanagi and K. Watanabe, ``Scrambling time from local
perturbations of the eternal BTZ black hole,'' JHEP 1508, 011 (2015) [arXiv:1503.08161
[hep-th]].
\item  S. H. Shenker and D. Stanford, ``Stringy effects in scrambling,'' JHEP 1505 (2015) 132
[arXiv:1412.6087 [hep-th]].
\item A. P. Reynolds and S. F. Ross, ``Butterflies with rotation and charge,''  [arXiv:1604.04099 [hep-th]].
\item T. Andrade, S. Fischetti, D, Marolf, S. F. Ross and M. Rozali, `` Entanglement and Correlations near Extremality CFTs dual to Reissner-Nordstrom $AdS_5$,'' JHEP 4 (2014) 23 [arXiv:1312.2839 [hep-th]].
\item N. Sircar, J. Sonnenschein and W. Tangarife, ``Extending the scope of holographic
mutual information and chaotic behavior,'' JHEP 1605 (2016) 091 [arXiv:1602.07307 [hep-th]]. 
\item S. Kundu and J. F. Pedraza, ``Aspects of Holographic Entanglement at Finite Temperature and Chemical Potential,'' JHEP 08 (2016) 177 [arXiv:1602.07353 [hep-th]] ; W. Fischler, A. Kundu, and S. Kundu, ``Holographic Mutual Information at Finite Temperature,'' PRD87 (2013)126012 [arXiv:1212.4764 [hep-th]].
\item M.A. Melvin, ``Pure magnetic and electric geons,'' Phys. Lett. 8 (1964) 65.
\item G. T. Horowitz and D. L. Welch, ``Duality invariance of the Hawking temperature and entropy,” Phys. Rev. D 49 (1994) 590 [arXiv: 9308077 [hep-th]];  Wung-Hong Huang, `` Entropy of Black-Branes System and T-Duality,”  Gen. Rel. Grav. 43 (2010) 1443 [arXiv: 0910.4633 [hep-th]]. 
\item W. Fischler, A. Kundu, and S. Kundu, ``Holographic Entanglement in a Noncommutative Gauge Theory,'' JHEP 01 (2014) 137 [arXiv:1307.2932 [hep-th]]..
\item F. Dowker, J. P. Gauntlett, D. A. Kastor and Jennie Traschen, ``Pair Creation of Dilaton Black Holes,'' Phys.Rev. D49 (1994) 2909-2917  [hep-th/9309075].
\item F. Dowker, J. P. Gauntlett, D. A. Kastor and J. Traschen, ``The decay of magnetic fields in Kaluza-Klein theory,'' Phys. Rev. D52 (1995) 6929 [hep-th/9507143]; M.~S.~Costa and M.~Gutperle, ``The Kaluza-Klein Melvin solution in M-theory,'' JHEP 0103 (2001) 027 [hep-th/0012072].
\item G.~W.~Gibbons and D.~L.~Wiltshire, ``Space-time as a membrane in higher dimensions,'' Nucl.\ Phys.\ B287 (1987) 717 [hep-th/0109093]; G.~W.~Gibbons and K.~Maeda, ``Black holes and membranes in higher dimensional theories with dilaton fields,'' Nucl.\ Phys.\ B298 (1988) 741; A. Chodos and S. Detweiler, Gen. Rel. and Grav. 14 (1982) 879; G.W. Gibbons and D.L. Wiltshire, Ann.Phys. 167 (1986) 201; erratum  ibid. 176 (1987) 393.
\item T. Friedmann and H. Verlinde, ``Schwinger pair creation of Kaluza-Klein particles: Pair creation without tunneling,'' Phys.Rev. D71 (2005) 064018 [hep-th/0212163]; L. Cornalba and M.S. Costa, ``A New Cosmological Scenario in String Theory,'' Phys.Rev. D66 (2002) 066001 [hep-th/0203031]; L. Cornalba, M.S. Costa and C. Kounnas, ``A Resolution of the Cosmological Singularity with Orientifolds,'' Nucl.Phys. B637 (2002) 378-394 [hep-th/0204261].
\item  M.J. Duff, H. L¨u and C.N. Pope, ``The Black Branes of M-theory,'' Phys.Lett. B382 (1996) 73 [hep-th/9604052].
\item I.R. Klebanov, A.A. Tseytlin, ``Entropy of Near-Extremal Black p-branes ,''  Nucl.Phys.B475 (1996) 164 [hep-th/9604089].
\item Wung-Hong Huang, `` Semiclassical Strings in Electric and Magnetic Fields Deformed $AdS_5 \times S^5$ Spacetimes,''  Phys.Rev. D73(2006) 026007 [hep-th/0512117].
\item A. Hashimoto and K. Thomas, ``Dualities, Twists, and Gauge Theories with Non-Constant Non-Commutativity,'' JHEP 01 (2005) 033, [hep-th/0410123].
\item P. Ginsparg and C. Vafa, Nucl. Phys. B289 (1987) 414; T. Buscher, Phys. Lett.
B159 (1985) 127; B194 (1987) 59; B201 (1988) 466; S. F. Hassan, ``T-Duality, Spacetime
Spinors and R-R Fields in Curved Backgrounds,'' Nucl.Phys. B568 (2000) 145
[hep-th/9907152].
\item  N. Seiberg and E. Witten, ``String Theory and Non-commutative Geometry,'' JHEP
09 (1999) 032 [hep-th/9908142].
\item   J. Bhattacharya, M. Nozaki, T. Takayanagi and T. Ugajin, ``Thermodynamical Property of Entanglement Entropy for Excited States,'' Phys. Rev. Lett. 110 (2013) 091602 [arXiv:1212.1164 [hep-th]].
\item   D. D. Blanco, H. Casini, L. Y. Hung and R. C. Myers, ``Relative Entropy and Holography,'' JHEP 1308 (2013) 060 [arXiv:1305.3182 [hep-th]].
\item  A. Hashimoto and K. Thomas, ``Non-commutative gauge theory on D-branes in Melvin universes,'' JHEP 01 (2006) 083  [arXiv:hep-th/0511197].
\item  D. Dhokarh, S. S. Haque and A. Hashimoto, ``Melvin Twists of global AdS(5) x S(5) and their
Non-Commutative Field Theory Dual'', JHEP 0808, 084 (2008) [ arxiv:0801.3812 [hep-th]].

\item  M. S. Costa, C. A. R. Herdeiro, and L. Cornalba, ``Flux-branes and the dielectric effect in string theory,'' Nucl. Phys. B619 (2001) 155–190  [arXiv: hep-th/0105023].
\item M. Gutperle and A. Strominger, ``Fluxbranes in string theory,'' JHEP 06 (2001) 035  [arXiv: hep-th/0104136].
\item A. Bergman and O. J. Ganor, ``Dipoles, twists and noncommutative gauge theory,''
JHEP 10 (2000) 018 [arXiv: hep-th/0008030].
\item A. Bergman, K. Dasgupta, O. J. Ganor, J. L. Karczmarek, and G. Rajesh, ``Nonlocal
field theories and their gravity duals,'' Phys. Rev. D65 (2002) 066005 [arXiv: hep-th/0103090].
\item O. J. Ganor and U. Varadarajan, ``Nonlocal effects on D-branes in plane-wave
backgrounds,'' JHEP 11 (2002) 051 [arXiv: hep-th/0210035].
\item O. Lunin and J. M. Maldacena, ``Deforming field theories with U(1) × U(1) global
symmetry and their gravity duals,'' JHEP 05 (2005) 033 [arXiv: hep-th/0502086].

\end{enumerate}
\end{document}